\documentclass[%
prb,
superscriptaddress,
rsi,%
 amsmath,amssymb,
preprint,%
]{revtex4-1}

\usepackage{graphicx}
\usepackage{dcolumn}
\usepackage{bm}
\usepackage{braket}
\usepackage{multirow}
\usepackage[dvipdfmx]{color}
\usepackage{color}
\usepackage[top=20truemm,bottom=20truemm,left=20truemm,right=20truemm]{geometry}

\newcount\num
\begin{document}


\title[Sample title]{Effect(s) of Cobalt Substitution in L1$_{\bm 0}$-(Fe,Co)Pt Thin Films}

\author{Shoya Sakamoto}
\affiliation{Department of Physics, The University of Tokyo, 7-3-1 Hongo, Bunkyo-ku, Tokyo 113-0033, Japan}

\author{Kumar Srinivasan}%
\author{Rui Zhang}%
\altaffiliation{contributed during the stay at Western Digital}
\author{Oleg Krupin}
\altaffiliation{contributed during the stay at Western Digital}
\affiliation{Western Digital Media, 1710 Automation Pkwy., San Jose, California 95131, USA}%

\author{Keisuke Ikeda}
\author{Goro Shibata}
\author{Yosuke Nonaka}
\author{Zhendong Chi}
\affiliation{Department of Physics, The University of Tokyo, 7-3-1 Hongo, Bunkyo-ku, Tokyo 113-0033, Japan}

\author{Masako Sakamaki}
\author{Kenta Amemiya}
\affiliation{Institute of Materials Structure Science, High Energy Accelerator Research Organization, 1-1 Oho, Tsukuba, Ibaraki 305-0801, Japan}

\author{Atsushi Fujimori}
\affiliation{Department of Physics, The University of Tokyo, 7-3-1 Hongo, Bunkyo-ku, Tokyo 113-0033, Japan}
\author{Antony Ajan}
\affiliation{Western Digital Media, 1710 Automation Pkwy., San Jose, California 95131, USA}%

\date{\today}

\begin{abstract}
We have studied the effect of cobalt substitution in $L1_{0}$-Fe$_{1-x}$Co$_{x}$Pt films by means of x-ray magnetic circular dichroism (XMCD) and first-principles calculations.
The magnetic moments of Fe ($\sim$2.5 $\mu_{\rm B}$) and Co ($\sim$1.5 $\mu_{\rm B}$) deduced using XMCD were almost unchanged upon Co doping, and the net magnetization decreases with increasing Co content.
Calculation also showed that the 3$d$ electrons that have been added by Co substitution occupy only spin-down bands. 
\end{abstract} 

\pacs{Valid PACS appear here}
\keywords{Suggested keywords}
\maketitle

%


Owing to the development of information technology, the amount of information has been increasing explosively. 
In such circumstances, hard disk drives (HDDs) have played a central role as storage devices, where the assembly of ferromagnetic grains is used as a bit. 
To achieve even higher areal bit density by decreasing the size of the grains, it is necessary to use materials with large magnetic anisotropy so that the decrease of the grain volume $V$ is compensated by the increase of the magnetic anisotropy energy $K$, maintaining the thermal stabilization factor $KV$/$k_{\rm B}T$ large enough, say above 60 \cite{Charap:1997aa}.
$L1_{0}$-ordered FePt and CoPt are promising materials since they possess very large $K$ of $7.0 \times 10^7$ erg/cc and $4.9 \times 10^7$ erg/cc, respectively \cite{Kryder:2008aa, Weller:2013aa}.   
To utilize these materials as a media for HDDs, a large value of saturation magnetization $M_{\rm S}$ is also required to increase the signal level and to achieve an adequate signal to noise ratio (SNR).
The reported $M_{\rm S}$ values of $L1_{0}$ FePt and CoPt are about 1100 emu/cc\cite{Okamoto:2002aa, Bublat:2010aa, Weller:2013aa} and 800 emu/cc \cite{Grange:2000aa, Weller:2013aa}, respectively, at room temperature. 

In previous studies employing macroscopic magnetometry such as SQUID or vibrating sample magnetometer (VSM) measurements, mixed compounds Fe$_{1-x}$Co$_{x}$Pt were reported to have larger net magnetization than pristine FePt, which would be a promising property for future HDD applications.
A 14\% increase was observed for (Fe$_{0.5}$Co$_{0.5}$)$_{60.5}$Pt$_{39.5}$ bulk specimen \cite{Saha:2002aa}, 15\% for (Fe$_{0.85}$Co$_{0.15}$)$_{50}$Pt$_{50}$ thin film \cite{Liu:2013aa}, and even 54\% for (Fe$_{0.8}$Co$_{0.2}$)$_{59}$Pt$_{41}$ thin film \cite{Lai:2007aa}.
However, first-principles calculations \cite{Maclaren:2001aa, Cuadrado:2014aa} suggested that the net magnetization decreases with increasing Co content.

In order to resolve this issue, we have performed x-ray absorption spectroscopy (XAS) and x-ray magnetic circular dichroism (XMCD) measurements, by which one can obtain the (effective) spin and orbital magnetic moments of each constituent element separately \cite{Carra:1993aa, Thole:1992aa, Stohr:1995ab}.
Since XAS and XMCD are also sensitive to chemical states \cite{Groot:1994aa}, it is possible to deduce the intrinsic magnetic moments excluding contributions from impurities such as surface oxides \cite{Takeda:2008aa, Sakamoto:2016aa}. In addition, we have done first-principles calculations to deduce the magnetic dipole term in the spin sum rule and to discuss the effect of Co substitution in FePt films. 

Samples were grown using the dc and rf sputtering methods on glass substrates. 
The sample structure was ``C cap (4 nm)/Fe$_{1-x}$Co$_{x}$Pt (22 nm)/MgO (20 nm)/seed/glass substrate" ($x=$ 0, 0.05, 0.1, 0.15, 0.3). Because we found strong oxide peaks in the XAS spectrum of the $x=0$ sample, we used another sample with the different structure of ``C cap (4 nm)/FePt (5 nm)/MgO (5 nm)/seed/glass substrate" for XAS and XMCD measurements.
The MgO, C layers were grown at room temperature, while Fe$_{1-x}$Co$_{x}$Pt layer was grown at the elevated temperature of 600 $^\circ$C to achieve the (001)-oriented $L1_{0}$-ordered phase \cite{Barmak:2005aa}. 
We used separate Fe, Co and Pt targets to control the composition.

XAS and XMCD measurements were conducted at beam line BL-16A1 of Photon Factory (PF), High Energy Accelerator Research Organization (KEK). 
The spectra were taken at room temperature in the total-electron-yield (TEY) mode.
A magnetic field of 5 T was applied parallel to the incident x rays and perpendicular to the film surface so that the magnetization of Fe$_{1-x}$Co$_{x}$Pt was fully saturated.
A double-step function with amplitude ratio 2:1 representing the $L_{3}$- and $L_{2}$-edge jumps has been subtracted from each absorption spectrum \cite{Chen:1995aa}.

First-principles calculations with the local-density approximation (LDA) were performed using a WIEN2k package. Spin-orbit interaction was also included. In order to study the effect of Co substitution, we constructed supercells containing Co atom at the Fe sites as shown in the supplemental material \footnote{see Supplemental Materials for further information}.\num=\value{footnote}
We assumed Vegard's relation of $a=3.863(1-x)+3.806x$ and $c=3.710(1-x)+3.684x$ for the lattice constants of Fe$_{1-x}$Co$_{x}$Pt \cite{Barmak:2004aa,Galanakis:2000aa}.

\begin{figure}
\begin{center}
\includegraphics[width=8.5 cm]{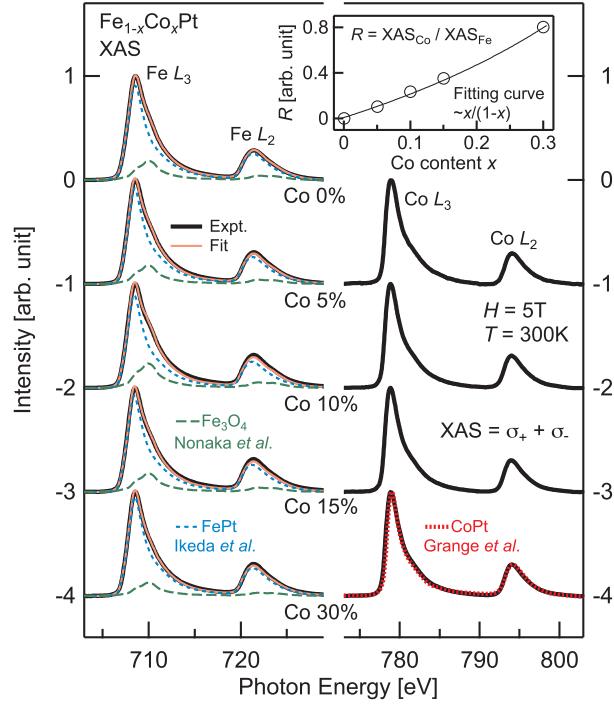}
\caption{Fe and Co $L_{2,3}$-edge XAS spectra of Fe$_{1-x}$Co$_{x}$Pt. 
The Fe XAS spectra were fitted by the summation of FePt \cite{Ikeda:2017aa} (blue dashed) and Fe$_{3}$O$_{4}$ (green dashed) \cite{Nonaka:2015aa} spectra. The fitted results are shown by pink solid curves. The XAS spectrum of $L1_{0}$-CoPt \cite{Grange:2000aa} is also shown by a red dashed curve as a reference. The inset shows the ratio $R$ of Co $L_{2,3}$ XAS integral to the Fe one as a function of Co content $x$.} 
\label{XAS}
\end{center}
\end{figure}

Figure \ref{XAS} shows the XAS spectra of Fe$_{1-x}$Co$_{x}$Pt recorded at the Fe and Co $L_{2,3}$ absorption edges, where they have been normalized to unity at the peak of the $L_{3}$ absorption edge for each element. 
The Fe XAS spectra show a shoulder around $\sim$710 eV, which did not in the previous studies of FePt films \cite{Dmitrieva:2007aa, Ikeda:2017aa}. This can be attributed to signals from Fe oxides formed at the surface because the position of the shoulder corresponds to the peak position of the XAS spectra of Fe$^{3+}$ compounds such as Fe$_{2}$O$_{3}$ \cite{Dmitrieva:2007aa} and because the intensities of the shoulder showed non-monotonic behavior as a function of Co content. 
Such shoulders were not seen in the Co XAS spectra, where the spectral line shape did not change upon Co doping. 
This can be naturally understood because Co is more difficult to be oxidized than Fe. 
Since the Fe XAS spectra contain signals from surface oxides, we have decomposed the spectra into FePt \cite{Ikeda:2017aa} and Fe oxides components by least-square fitting. 
Here, we have used the Fe$_{3}$O$_{4}$ spectrum \cite{Nonaka:2015aa} in order to reflect the contributions from both Fe$^{2+}$ and Fe$^{3+}$ oxides.
Note that the following discussion would not be affected if we adopt the Fe$_{2}$O$_{3}$ spectrum, but fit using the Fe$_{3}$O$_{4}$ better reproduced the experimental spectra as was also mentioned in Ref. \onlinecite{Dmitrieva:2007aa}.
The fitted results are shown by pink solid curves, and the FePt and Fe$_{3}$O$_{4}$ spectrum components are also separately shown by blue and green dashed curves, respectively.
This decomposition is crucial when estimating the magnetic moments using XMCD sum rules \cite{Carra:1993aa,Thole:1992aa}.

The inset of Fig. \ref{XAS} shows the ratio of the XAS area of Co to that of Fe as a function of Co content $x$. The data are well fitted to $Ax/(1-x)$, where $A$ is a proportionality constant. This confirms that the actual concentration of Fe and Co atoms, or at least its ratio, is very close to the designed values. 


\begin{figure}
\begin{center}
\includegraphics[width=8.5 cm]{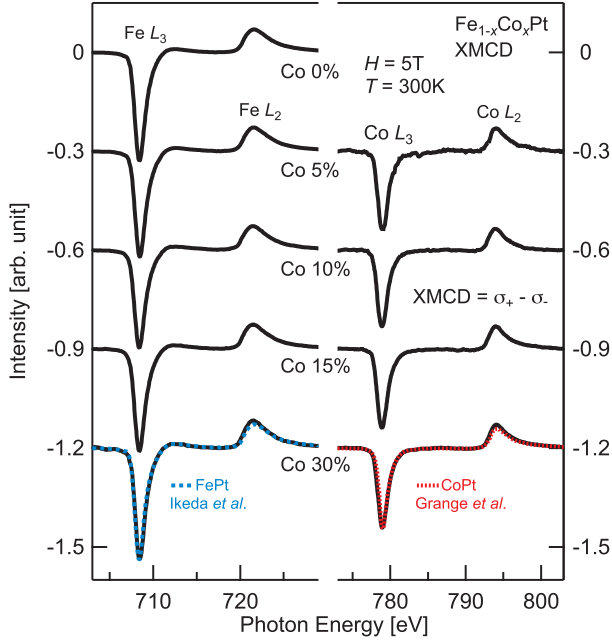}
\caption{XMCD spectra of Fe$_{1-x}$Co$_{x}$Pt at Fe and Co $L_{2,3}$ edges. The spectra of $L1_{0}$-ordered FePt and CoPt are shown as references by blue and red dashed curves, respectively.}
\label{XMCD}
\end{center}
\end{figure}

Figure \ref{XMCD} shows the XMCD spectra of Fe$_{1-x}$Co$_{x}$Pt. They are nearly identical to those of FePt \cite{Ikeda:2017aa} and CoPt \cite{Grange:2000aa} as shown by blue and red dashed curves, respectively. This indicates that the local electronic structure of the Fe$_{1-x}$Co$_{x}$Pt alloy is similar to their two end members FePt and CoPt. Note that the Fe-oxide shoulder seen in the XAS spectra was absent in the XMCD spectra, presumably because naturally formed Fe oxides are usually non-ferromagnetic.


\begin{table*}
\begin{center}
\caption{Orbital and (effective) spin magnetic moments of Fe and Co in Fe$_{1-x}$Co$_{x}$Pt. $m_{l}$ and $m_{s}$ denote the orbital and spin magnetic moments, respectively, and $m_{s}^{\rm eff} = m_{s}+7m_{T}$ the effective spin magnetic moment, where $m_{T}$ is the magnetic dipole term. $m_{\rm Fe}$ and $m_{\rm Co}$ represent the summation of the orbital and spin magnetic moments for each atom, where $m_{T}$ term has been subtracted as described in the main text. $m_{\rm net}$ ($= (1-x)m_{\rm Fe}+xm_{\rm Co}$) represents the net magnetic moment, and $M_{\rm net}$ the net magnetization of the film in the units of emu/cc taking the Pt magnetic moments of $0.38(1-x)+0.42x$ $\mu_{\rm B}$ into account.
Note that uncertainties are shown in parentheses.}
\begin{tabular}{p{5.4em}p{3.1em}p{2.1em}p{2.2em}p{3.5em}p{3.6em}p{0.1em}p{3.1em}p{2.2em}p{2.3em}p{3.5em}p{3.6em}p{0.0em}p{4.0em}p{3.8em}}

\hline\hline 
\multirow{2}{*}{${\rm Fe}_{1-x}{\rm Co}_{x}{\rm Pt}$} & \multicolumn{5}{c}{Fe [$\mu_{\rm B}$]} & & \multicolumn{5}{c}{Co [$\mu_{\rm B}$]} & & \hfil $m_{\rm net}$ \hfil & \hfil$M_{\rm net}$\hfil\\ 
 \cline{2-6}\cline{8-12}
  & \hfil$m_{l}$\hfil &  \multicolumn{2}{c}{$m_{s}+7m_{T}$} & \hfil $m_{l}/m_{s}^{\rm eff}$ \hfil &\hfil $m_{\rm Fe}$\hfil & &  \hfil$m_{l}$\hfil & \multicolumn{2}{c}{$m_{s}+7m_{T}$} & \hfil $m_{l}/m_{s}^{\rm eff}$ \hfil & \hfil $m_{\rm Co}$\hfil & & \hfil [$\mu_{\rm B}$] \hfil & \hfil [emu/cc]\hfil\\
\hline

$x=0$ &  \hfil0.27(3)\hfil & \multicolumn{2}{c}{2.07(10)} &\hfil0.13(2) \hfil & \hfil2.49(11)\hfil& & \hfil-\hfil & \multicolumn{2}{c}{-}&\hfil-\hfil & \hfil-\hfil & &\hfil 2.49(11) \hfil &\hfil 961 \hfil\\ 
$0.05$ &  \hfil0.25(3)\hfil & \multicolumn{2}{c}{2.06(10)} &\hfil0.12(1) \hfil &\hfil2.47(11)\hfil& & \hfil0.22(4)\hfil & \multicolumn{2}{c}{1.25(13)}& \hfil0.17(3) \hfil & \hfil1.45(13)\hfil & & \hfil2.42(10)\hfil &\hfil 942 \hfil\\
$0.10$ &  \hfil0.19(2)\hfil & \multicolumn{2}{c}{2.04(10)} &\hfil0.10(1) \hfil & \hfil2.39(10)\hfil& & \hfil0.23(4)\hfil & \multicolumn{2}{c}{1.12(12)}&\hfil0.20(4) \hfil & \hfil1.33(11)\hfil & & \hfil2.28(9)\hfil &\hfil 896\hfil\\
$0.15$ &  \hfil0.20(2)\hfil & \multicolumn{2}{c}{2.02(10)} &\hfil0.10(1) \hfil &\hfil2.36(10)\hfil& & \hfil0.22(4)\hfil & \multicolumn{2}{c}{1.27(13)}&\hfil0.17(3) \hfil & \hfil1.46(12)\hfil &  & \hfil2.23(9)\hfil &\hfil 881 \hfil\\
$0.30$ &  \hfil0.20(2)\hfil & \multicolumn{2}{c}{2.12(11)} &\hfil0.10(1) \hfil &\hfil2.45(11)\hfil& & \hfil0.17(3)\hfil & \multicolumn{2}{c}{1.31(14)}&\hfil0.13(3) \hfil & \hfil1.47(12)\hfil & & \hfil 2.15(8) \hfil &\hfil 863 \hfil\\

0 (Expt.\cite{Dmitrieva:2007aa})& \hfil0.19\hfil &\multicolumn{2}{c}{2.21}& \hfil0.09\hfil &\hfil$^{\ }2.40^{\ast}$\hfil & & \hfil-\hfil &\multicolumn{2}{c}{-}& \hfil-\hfil & \hfil-\hfil & & \hfil$^{\ }2.40^{\ast}$\hfil& \hfil $^{\ }931^{\ast}$ \hfil\\
0 (LSDA)& \hfil0.072\hfil & \hfil2.87\hfil & \hfil-0.25\hfil &\hfil0.025\hfil &\hfil2.94\hfil & & \hfil-\hfil & \multicolumn{2}{c}{-}&\hfil-\hfil & \hfil-\hfil & & \hfil2.94\hfil & \hfil 1110 \hfil\\
1 (Expt.\cite{Grange:2000aa})& \hfil-\hfil & \multicolumn{2}{c}{-}& \hfil-\hfil &\hfil-\hfil & & \hfil0.26\hfil & \multicolumn{2}{c}{1.76} & \hfil0.148\hfil & \hfil$^{\ }2.02^{\ast}$\hfil & & \hfil$^{\ }2.02^{\ast}$\hfil & \hfil $^{\ }840^{\ast}$ \hfil\\
1 (LSDA)& \hfil-\hfil & \multicolumn{2}{c}{-} & \hfil-\hfil &\hfil-\hfil & & \hfil0.109\hfil & \hfil1.79\hfil & \hfil0.033\hfil &\hfil0.061\hfil &\hfil1.90\hfil & & \hfil1.90\hfil & \hfil 806 \hfil\\
0-1 (CPA\cite{Maclaren:2001aa})& \hfil-\hfil & \multicolumn{2}{c}{-} &\hfil-\hfil &\hfil2.9\hfil & & \hfil-\hfil & \multicolumn{2}{c}{-} &\hfil-\hfil & \hfil1.8\hfil & & \hfil$2.9-1.1x$\hfil\\
\hline\hline
\multicolumn{5}{l}{$^{\ast}$Effective values containing 7$m_{T}$ term.}
\end{tabular}
\label{table}
\end{center}
\vspace{0cm}
\end{table*}

We have estimated the spin and orbital magnetic moments of Fe and Co by applying the XMCD sum rules \cite{Chen:1995aa}:
\begin{eqnarray}
&\displaystyle m_{l}=-\frac{4\int_{L_{2,3}}{\sigma_{+}-\sigma_{-}}\ {\rm d}\omega}{3\int_{L_{2,3}}{\sigma_{+}+\sigma_{-}}\ {\rm d}\omega}n_{h},\\
&\displaystyle m_{s}^{\rm eff}=-\frac{2\int_{L_{3}}{\sigma_{+}-\sigma_{-}}\ {\rm d}\omega-4\int_{L_{2}}{\sigma_{+}-\sigma_{-}}\ \rm{d}\omega}{\int_{L_{2,3}}{\sigma_{+}+\sigma_{-}}\ \rm{d}\omega}n_{h},\\
&\displaystyle m_{s}^{\rm eff}=m_{s}+7m_{T},
\end{eqnarray}
where $\sigma_{+}$ and $\sigma_{-}$ denote absorption cross sections, namely XAS intensity, measured with x rays with positive and negative helicity. $m_{l}$ and $m_{s}$ are the orbital and spin magnetic moments, $m_{s}^{\rm eff}$ the effective spin magnetic moment, and $m_{T}$ the magnetic dipole term in units of $\mu_{\rm B}$/atom. Note that each term can be written as $m_{s} =  -2 \mu_{\rm B} \Braket{S_{\alpha}}/\hbar$, $m_{l} =  - \mu_{\rm B} \Braket{L_{\alpha}}/\hbar$, and $m_{T} =  - \mu_{\rm B} \Braket{T_{\alpha}}/\hbar$, where ${\bm S}$ denotes the spin angular momentum operator, ${ \bm L}$ the orbital angular momentum operator ${\bm T}$ the magnetic dipole operater ${\bm T}={\bm S}-3\hat{r}(\hat{r}\cdot{\bm S})$, and $\alpha$ represents the incident x-ray direction \cite{Stohr:1995aa}. The number of holes in the Fe and Co 3$d$ shell, denoted by $n_{h}$, was assumed to be 3.4 and 2.63, respectively \cite{Dmitrieva:2007aa, Grange:2000aa}. As for the Fe XAS area, which appears in the denominators of Eqs. (1) and (2), we used the fitted FePt component to exclude oxide contributions. By applying this correction, the estimated magnetic moments increased by at most $\sim$30\%.


Although $m_{T}$ term can be ignored in cubic systems with $T_{d}$ or $O_{h}$ symmetry, it cannot be neglected in the case of highly anisotropic systems such as $L1_{0}$-ordered alloys. Here we estimate the $m_{T}$ of Fe (Co) atom in FePt (CoPt) as follows.
When spin-orbit coupling is weak and charge and spin are decoupled as in 3$d$ transition metals, $m_{T}$ can be expressed as \cite{Stohr:1995aa}
\begin{eqnarray}
&\displaystyle 7m_{T} \approx \sum_{i}\frac{7}{2}Q_{\alpha}^{i}m_{s}^{i} = \sum_{i}\frac{7}{2}Q_{\alpha}^{i}\Delta n_{i}\mu_{\rm B},\\
&\displaystyle Q_{\alpha}^{i} = \langle d_{i}|Q_{\alpha \alpha}|d_{i}\rangle, Q_{\alpha \beta}=\delta_{\alpha\beta} - 3r_{\alpha}r_{\beta}/r^2,
\end{eqnarray}
where $\alpha, \beta=x,y$, or $z$ denotes the Cartesian frame. $m_{s}^{i}$ is the spin magnetic moment of electrons in a $d_{i}$ orbital, which is directly related to $\Delta n_{i}$ (=$n_{i, \uparrow}-n_{i, \downarrow}$) the difference between occupation numbers of spin-up and spin-down electrons in the orbital.
The values of $Q_{\alpha}^{i}$ are summarized in Ref. \onlinecite{Stohr:1995aa}, and $\Delta n_{i}$ can be easily calculated by first-principles calculation. In the out-of-plane geometry, $7m_{T}$ of FePt and CoPt were calculated to be -0.246 $\mu_{\rm B}$ and 0.0326 $\mu_{\rm B}$, respectively, while in the in-plane geometry, 0.123 $\mu_{\rm B}$ and -0.0171 $\mu_{\rm B}$ \footnote{In the out-of-plane geometry, where the direction of the light $\vec{k}$ and the applied magnetic field $\vec{H}$ are parallel to the $z$ axis of the lattice or to the surface normal, $7m_{T}^{z}/\mu_{\rm B} \approx -2\Delta n_{z^2} +2 \Delta n_{xy}+2 \Delta n_{x^{2}-y^{2}}-\Delta n_{yz}-\Delta n_{xz}$. In the in-plane geometry, where $\vec{k}$, $\vec{H} \parallel \vec{x}$, $7m_{T}^{x}/\mu_{\rm B} \approx \Delta n_{z^2} - \Delta n_{xy}- \Delta n_{x^{2}-y^{2}}+2\Delta n_{yz}-\Delta n_{xz}$ or just $m_{T}^{x}=-m_{T}^{z}/2$.}.
These values agree with the sum rules applications to the calculated XMCD spectra by first-principles calculations \cite{Galanakis:2002aa}. Therefore, one should be careful about the spin magnetic moment of Fe in FePt with high degree of $L1_{0}$-order, where $m_{s}^{\rm eff}=m_{s}+7m_{T}$ may be smaller than $m_{s}$ by at most $\sim$15\%. 
Note that we could confirm this simple calculation reproduced the reported values of $m_{T}$ of CrO$_{2}$ \cite{Komelj:2004aa}. 

The obtained effective spin and orbital moments are summarized in TABLE \ref{table} together with the previous experimental and theoretical results. The obtained spin magnetic moments of Fe are similar to the previous experimental study \cite{Dmitrieva:2007aa}, but those of Co are smaller than the results of Ref. \onlinecite{Grange:2000aa}.
This is probably because some of the Co atoms are magnetically inactive near the surface with oxidized Fe. In fact, if the area ratios of the raw Fe XAS spectra to the fitted FePt components are also multiplied to the Co magnetic moments, the values become as large as those in Ref. \onlinecite{Grange:2000aa}. 

\begin{figure}
\begin{center}
\includegraphics[width=8.4 cm]{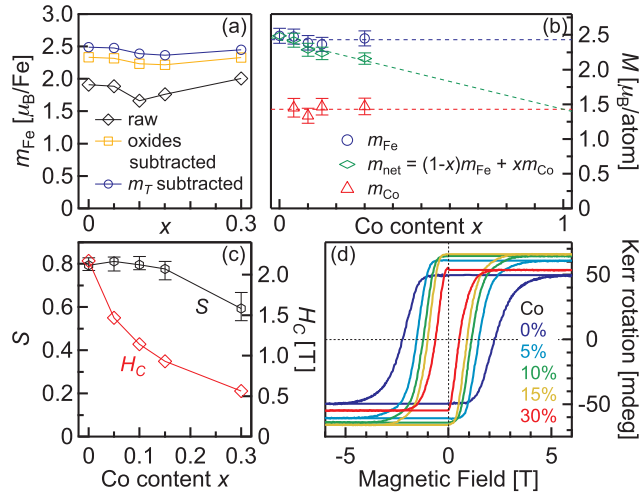}
\caption{(a) Magnetic moments of Fe before and after the corrections. (b) Magnetic moments of Fe and Co, and the net magnetic moment of Fe$_{1-x}$Co$_{x}$Pt as a function of Co content. (c) Degree of $L1_{0}$ order and coercive field. (d) Out-of-plane magnetic hysteresis measured by magneto-optical Kerr effect.}
\label{moment}
\end{center}
\end{figure}

Figure \ref{moment}(a) shows the magnetic moments $m_{s}+m_{l}$ of Fe taking the oxides component and $m_{T}$ term into account. 
Although the raw data (black diamonds) showed non-monotonic behavior, the moments became almost constant after the subtraction of the oxides component from the raw XAS spectra as shown by orange squares.
Further, the subtraction of the $m_{T}$ term had a minor effect on how the magnetic moments behaves with Co content, as shown by blue circles. Here, the $7m_{T}$ term was most simply estimated as $(7m_{T} / m_{s})^{\rm calc} m_{s}^{\rm expt} S$ by assuming a linear relationship between $m_{T}$ and the degree of $L1_{0}$ order $S$. 
$S$ was deduced from the intensity ratio of the superlattice peak $I(110)$ to the fundamental peak $I(200)$ in the in-plane XRD profiles \footnote{Because the in-plane orientation of $a$ and $b$ axes are random in our granular films, one could directly compare the in-plane XRD intensity with simulated powder diffraction profile.} as $S^2 = \frac{(I(110)/I(200))^{\rm obs}}{(I(110)/I(200))^{\rm theor}}$ \cite{Yang:2012aa} and is plotted in Fig. \ref{moment}(c). 
Note that the $m_{T}$ term for Co is negligibly small according to our LDA calculation. 

Figure \ref{moment}(b) summarizes the corrected magnetic moments of Fe and Co and the net magnetic moments of Fe$_{1-x}$Co$_{x}$Pt. 
Since both the Fe and Co magnetic moments were almost constant as functions of Co content, the net magnetic moment was found to decrease with Co doping. This disagrees with the previous experiments \cite{Saha:2002aa, Liu:2013aa, Lai:2007aa} but agrees with the theoretical prediction \cite{Maclaren:2001aa}. In Fig. \ref{moment}(b), blue and red solid lines represent averaged values of the magnetic moment of Fe and Co, respectively, and green one values calculated from them. Fitting using a linear function also reproduced the blue and green lines, although fitting to the Co magnetic moment showed slightly increasing behavior due to the fewer data points.
Note that the Pt magnetic moments, which can be as large as 0.38-0.42 $\mu_{\rm B}$ \cite{Galanakis:2002aa}, have not been considered.



Out-of-plane magnetic hysteresis curves were also measured by magneto-optical Kerr effect (MOKE) and are shown in Fig. \ref{moment}(d). Coercive field ($H_{C}$) is also plotted in Fig. \ref{moment}(c). In spite of the small change in the degree of $L1_{0}$ order caused by Co substitution, $H_{C}$ or the magnetic anisotropy decreases rather rapidly.

\begin{figure}[t!]
\begin{center}
\includegraphics[width=7.2 cm]{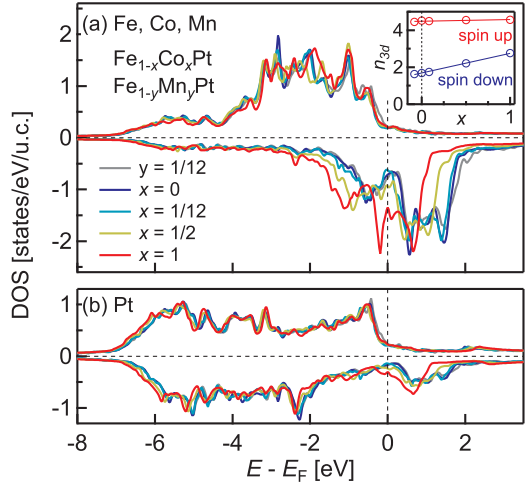}
\caption{Densities of states (DOSs) of Fe$_{1-x}$Co$_{x}$Pt and Fe$_{1-y}$Mn$_{y}$Pt. The partial DOSs of the 3$d$ transition metals and Pt are separately shown in panels (a) and (b). Inset shows the 3$d$ orbital occupation number each of the spin-up and spin-down bands.}
\label{DOS}
\end{center}
\end{figure}

In order to understand the microscopic mechanism why the magnetization decreased with Co doping, we have calculated the densities of states (DOSs) of Fe$_{1-x}$Co$_{x}$Pt.
Fig. \ref{DOS}(a) shows the Fe and Co partial DOS for varying the Co content. The DOSs of spin-up bands are almost fully occupied and unchanged upon doping, while those of spin-down bands shift upward without a significant change in the line shapes and concomitantly their occupation number increases. This suggests that the Co substitution does not change the band structure significantly, but simply provides the spin-down bands with electrons.
The inset shows the electron occupation numbers of the spin-up and spin-down 3$d$ bands ($n_{3d}$). The spin-up bands are almost completely filled with 4.5 electrons and the occupation number is nearly constant with Co content. 
On the other hand, the occupation number of the spin-down bands shows a linear increase.  
This can be understood because Co has one more 3$d$ electron than Fe, and further electron doping leads to the occupation of spin-down bands. 
Note that the Pt partial DOS showed little change as shown in Fig. \ref{DOS}(b), and occupation number of Pt 5$d$ orbitals was almost unchanged.

To summarize, the XMCD results have clarified that the net magnetic moment of Fe$_{1-x}$Co$_{x}$Pt decreases with increasing Co content because the additional electron provided by Co occupy the spin-down bands instead of the almost completely filled spin-up bands. 
The discrepancy between the present results and the previous experiments \cite{Saha:2002aa, Liu:2013aa, Lai:2007aa} would be attributed to some experimental artefacts such as errors in estimating the volume of FePt.  
Especially in the case of granular thin films, it is difficult to estimate the film thickness and the packing density of grains with sufficient accuracy. 
The present results also indicate that the magnetic moment may increase when holes are doped to the system. 
To elaborate this conjecture, we have also calculated the DOS of Mn-substitute FePt. The spin-down DOS indeed shows a shift to the opposite direction, i.e., upward as expected. The occupation number is also plotted in the inset at the position of $x=-1/12$, representing a hole doping, and extrapolates from the trend of electron doping by Co.
Therefore, the small amount of hole doping by Mn substitution or some other methods might be a better way to improve the magnetic properties of FePt. 
Actually, the effect of band filling\cite{Daalderop:1991aa,Sakuma:1994aa} and Mn substitution\cite{Burkert:2005aa} to the magnetic properties of FePt were theoretically studied, and it was reported that the magnetic anisotropy as well as magnetization can be enhanced.
However, too much doping may not be effective\cite{Meyer:2006aa} because $L1_{0}$-MnPt is antiferromagnetic.

This work was supported by Grants-in-Aid for Scientific Research from the JSPS (Grants No. 15H02109 and No. 15K17696).
The experiment was done under the approval of the Photon Factory Program Advisory Committee (Proposals No. 2013S2-004, 2016S2-005 and No. 2016G006).
S.S. acknowledges financial support from Advanced Leading Graduate Course for Photon Science (ALPS) and the JSPS Research Fellowship for Young Scientists. Z.C. acknowledges financial support from Materials Education program for the futures leaders in Research, Industry and Technology (MERIT).
A.F. is an adjunct member of Center for Spintronics Research Network (CSRN), the University of Tokyo, under Spintronics Research Network of Japan (Spin-RNJ).

\bibliography{BibTex_FeCoPt.bib}

\newpage

\part*{Supplementary Information}
\section*{Supercells}

\begin{figure}[h!]
\begin{center}
\includegraphics[width=16 cm]{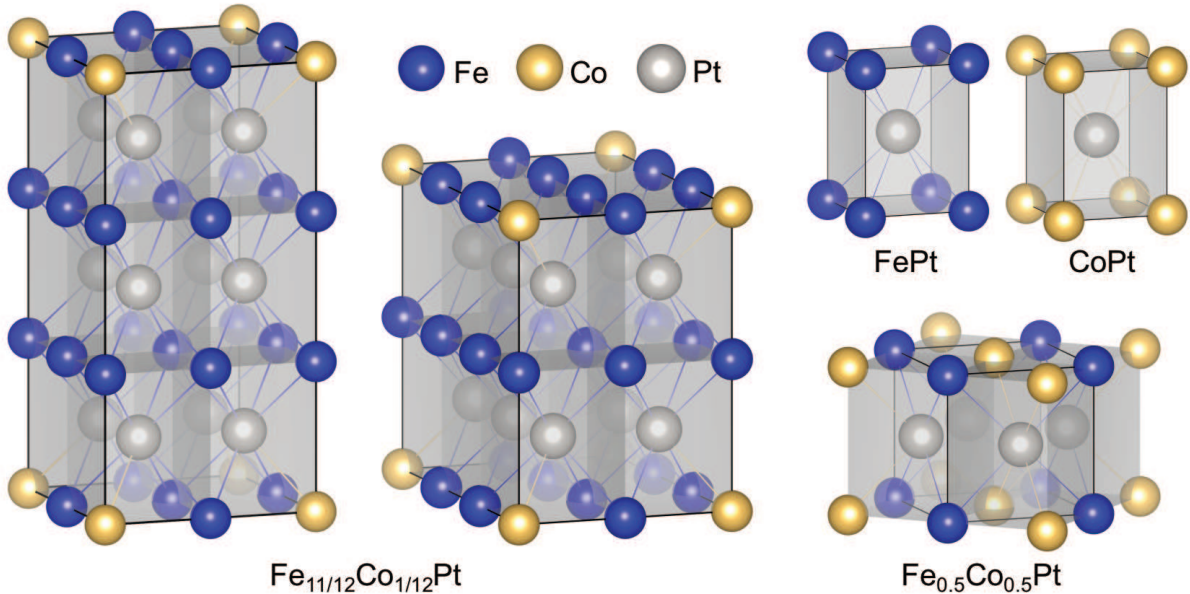}
\caption{Supercells of Fe$_{1-x}$Co$_{x}$Pt ($x$ = 0, 1/12, 0.5, 1) used in the first-principles calculations. }
\label{supercell}
\end{center}
\end{figure}

Figure \ref{supercell} shows the supercells of $L1_{0}$-Fe$_{1-x}$Co$_{x}$Pt ($x$ = 0, 1/12, 0.5, 1) used in the present first-principles calculation. We assumed the smallest or simplest possible supercells with periodically located Co atoms for each Co concentration.
For $x = 1/12$, two different supercells $2\times2\times3$ and $2\times3\times2$ are possible, but only the density of states (DOS) of the $2\times2\times3$ supercell is shown in the main manuscript.
Note that we have confirmed the DOS of the $2\times3\times2$ was almost identical to that of $2\times2\times3$, indicating that the effect of Co doping is just an electron doping to the spin-down bands.

\end{document}